\documentclass[a4paper,12pt]{article}
\usepackage{amsfonts}
\usepackage{amsmath}
\usepackage{amstext}
\usepackage{amsthm}
\usepackage{xspace}
\usepackage[dvips]{graphicx}

\newcommand\ac{\`a\xspace}

\newtheorem{theorem*}{Theorem}
\newtheorem{prop*} {Proposition}
\newtheorem{lemma*}{Lemma}

\theoremstyle{definition}
\newtheorem{definition*}{Definition}
\newtheorem{cor*}{Corollary}
\newtheorem{rem*}{Remark}
\theoremstyle{remark}

\newtheorem{dim*}{\bf Dimostrazione }

\newtheorem{guess*}{\bf Osservazione}

\input psfig.sty
\begin{document}
\title{A thermodynamical approach to dissipation range turbulence} 
\author{Jacopo Bellazzini}
\date{}
\maketitle
\begin{center}
\small  Dipartimento di Ingegneria Aerospaziale \\
\small Universit\ac di Pisa\\
 \small Via G. Caruso 56100 PISA -Italy\\ 
\end{center}

\begin{abstract}
A model to explain the statistics of the velocity gradients in the dissipation 
range of a turbulent flow is presented. 
The experimentally observed non-gaussian statistics 
is theoretically  predicted by means of a thermodynamical analogy using 
the maximum entropy principle of ordinary statistical mechanics.
\end{abstract}

\\
The dynamics of fluids at high Reynolds number is one of the most
interesting subject of statistical physics. In the limit of infinite
Reynolds number, the celebrated theory of Kolmogorov (K41) \cite{Ko} 
suggests that the small
scales statistics is characterized only by the mean rate of energy dissipation
per unit mass $\epsilon$ and the scale $l$. 
The K41 theory is based upon the concept of self-similarity
of the inertial range $[l_d,L]$ and implies that the velocity gradients
$\delta v_{l}(x)=|v(x+l)-v(x)|$ scale as
\begin{equation}
\label{scaling}
<\delta v_{l}^{q}> \sim (\epsilon l)^{q/3}.
\end{equation} 

The experimental evidence of the breakdown of eq.(\ref{scaling}), 
see \cite{Ba,An,Vi}, 
due to the presence of intermittency, induced Kolmogorov and Obukhov \cite{Ko2,Ob} to
modify the K41 theory introducing the fluctuation of the mean flux of energy 
$\epsilon$.
Indeed, as pointed out by Siggia \cite{Sg}, the turbulent flow 
can be described as the superposition of 
two fluid states, the coherent
structures and the random fluctuations. Hence, in the inertial range, the flow preserves  
organized structures as, for instance, vortex sheets and filaments.
The coherent structures are ``rare'' events 
that contribute to the non-gaussian high-amplitude fluctuations.
Many models for such superposition have been proposed, either based on a  
mapping-closure theory or on the fractal approach \cite{Kr,Fr,Benzi,Me,She, She2}.\\
However, also at very small scales $l \sim l_d$, when the coherent structures
are annihilated by dissipation, 
the tails of the velocity gradients statistics  are deeply non 
gaussian. Indeed, experimental results show that the tails of the 
probability density
function (PDF) of the velocity gradients are close to 
exponential \cite{Wa} for small $l$.
In this paper, we conjecture that the PDF of the velocity gradients in the
dissipation range, $l \sim l_d$, can be described by means of 
ordinary statistical mechanics. 
We propose to consider the velocity gradients $\delta v_l$ of the random 
incoherent flow 
like the momentum ${\bf p}$ of a particle in a ideal gas with a ``temperature''
$T$ that changes slowly. Specifically, we consider this temperature not 
to be fixed but 
$\chi^{2}$-distributed with degree $n=3$; the resulting 
PDF agrees both
with the experimental results and with the theoretical structure function.\\
In the ordinary  description of an ideal gas, the momentum ${\bf p}$ of a 
particle 
with mass $m$ may be considered as a random variable. From the maximum entropy 
principle one obtains the density distribution
\begin{equation}
\label{gas}
\psi({\bf p})=\frac{1}{Z}e^{-\beta\frac{{\bf p}^{2}}{2m}},
\end{equation} 
where $Z$ is the partition function and $\beta=\frac{1}{kT}$.
Here $T$ is the temperature of the thermal bath and $k$ is Boltzmann's 
constant. \\
Now, we consider the
temperature of the termal bath not to be constant in time, but 
fluctuating with a given distribution $f(T)$.
In practice, we suppose that the time scale on which $T$ fluctuates is much 
larger then the typical time to reach equilibrium. Let $\psi({\bf p}|T)$ be 
the conditional probability, and let $\psi({\bf p})$ be the probability to 
observe a 
certain value of ${\bf p}$ no matter what $T$ is. 
Then, the following relation holds
\begin{equation}
\label{prob}
\psi({\bf p})=\int_{0}^{+\infty}\psi({\bf p}|T)f(T)dT.
\end{equation} 
Due to the maximum entropy principle of ordinary statistical mechanics,
the conditional probability  $\psi({\bf p}|T)$ is gaussian but 
the probability to observe a certain value of ${\bf p}$ is not.\\
The main idea of the paper is to consider the velocity gradients $\delta v_{l}(x)$ in the dissipation range and in the limit of infinite Reynolds number, 
as the momentum ${\bf p}$ of a particle fulfilling 
eq.(\ref{gas}) with $m=k=1$. 
In our approach, the ``temperature'' corresponds to the averaged kinetic
energy fluctuation $<\delta v_l^{2}>$ and will be referred with the symbol
$T_t$. \\
As shown by Kraichnan \cite{Kra}, if we observe the statistics of the 
velocity gradients 
in the inertial range, considering
both the coherent structures and the random fluctuations, the 
equipartition of energy does not apply. However, in the dissipation range, if 
we suppose that the fluctuations of the velocity gradients are given 
only by the incoherent 
fluctuations, we argue (and assume) that the 
equipartition of energy holds.
The physical implication of this model
is that the velocity fluctuations, in the dissipation range, may be described 
as a cascade of thermal baths. 
At every scale the ordinary
statistical mechanics holds, i.e. the conditional probability for a fixed 
temperature $T_t$ is gaussian, but, due to the changes of such
temperature, the final distribution will be different from the gaussian one.\\
Due to the three dimensional nature of the turbulent flow, we suggest
that the temperature $T_t$  fluctuates with a
$ \chi^{2}$- distribution with degree $n=3$.
Actually, such physical model for the distribution of the averaged kinetic 
energy 
fluctuations $f(T_t)$ has been proposed recently by Beck \cite{Beck} in the
formalism of nonextensive statistical mechanics.\\
The temperature $T_t$ has to be considered as a random variable
given by the expression
\begin{equation}
T_t=\sum_{i=1}^{3} X_{i}^{2},
\end{equation}
where $X_{i}$ are independent gaussian variables with zero average.
The mean value of the averaged kinetic energy fluctuations $T_t$ 
depends on the 
scale $l$. In particular, at  scale 
$l$, the mean value $<T_t>$ is $3g(l)$, where $g(l)=<X^{2}>$. 
This means that the probability density $f(T_t)$ is given by
\begin{equation}
\label{chi}
f(T_t)=\frac{1}{\Gamma (3/2)}(\frac{1}{2 g(l)})^{3/2}T_t^{1/2}e^{-\frac{T_t}{2 g(l)}}.
\end{equation} 

\begin{figure}[h]
\begin{center}
\includegraphics[width=10cm]{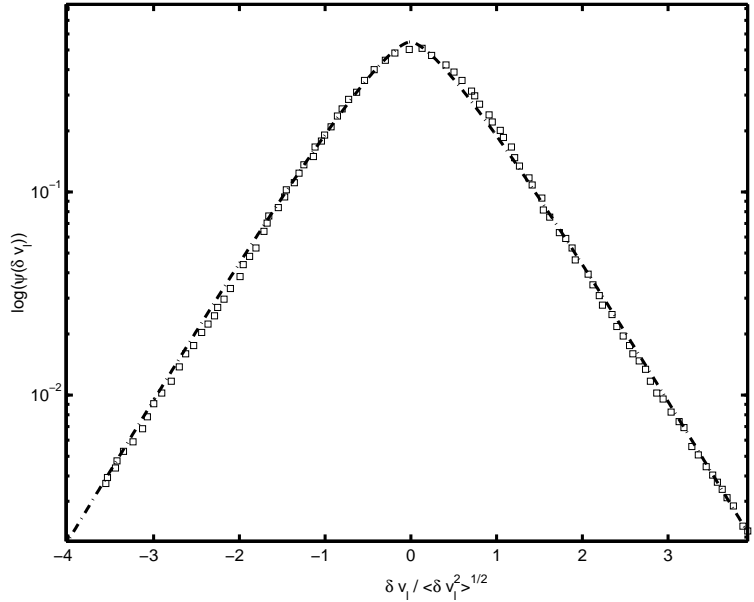}
\caption{\it Log-linear plot of the probability distributions $\psi (\delta v_l)$ 
of the velocity gradients. The theoretical prediction is obtained with l=0.58 (dashed-dotted line). The
experimental data of Ref.\cite{Wa} are indicated by symbols. }
\label{dist}
\end{center}\end{figure}

If we introduce eq.(\ref{chi}) into eq.(\ref{prob}) we obtain the following integral
\begin{equation}
\label{bessel}
\psi(\delta v_{l})=\frac{C}{g(l)^{3/2}}\int_{0}^{+\infty} e^{-\frac{\delta v_{l}^{2}}{2T_t}}e^{-\frac{T_t}{2g(l)}}dT_t,
\end{equation}
where $C=(4\sqrt{\pi}\Gamma (3/2))^{-1}$ is the normalization constant.
The latter integral can be evaluated
exactly. Indeed, eq.(\ref{bessel}) can be rewritten as
\begin{equation}
\label{risultato}
\psi(\delta v_{l})=\frac{2C}{g(l)}\delta v_l K_1\left(\frac{\delta v_l}{g(l)^{1/2}} \right),
\end{equation}  
where $K_1(\cdot)$ is the modified  Bessel function of the second kind.\\
Eq.(\ref{risultato}) provides a theoretical prediction of the statistics of 
the velocity gradients
 of the incoherent turbulence
in the limit of infinite Reynolds number. 
Obviously, the statistics of eq.(\ref{risultato}) depends on the scaling law 
$g(l)$ of the mean temperature $<T_t>$. \\
Given the relation between the mean temperature $<T_t>$ and the scale $l$, the 
moments of the velocity gradients can be evaluated easily by means of 
eq.(\ref{risultato}). Indeed, we obtain
\begin{equation}
<\delta v_l ^q> = f(q)g(l)^{q/2},
\label{mom}
\end{equation}
where $f(q)$ is a function of $q$ depending on the moments of the Bessel 
function $K_1(\cdot)$
\begin{equation}
f(q) = 2C \int y^{q+1}K_{1}(y)dy.
\end{equation} 
Equation (\ref{mom}) shows that the moments of the velocity
gradients depend only on the scaling law of the mean temperature $<T_t>$.\\
At the very small scales, $l \sim l_d$, the velocity field is smooth due to the
lack of singularities. Therefore, the structure function scales as 
$<\delta v_l ^q> \sim l^q$. Then, due to eq.(\ref{mom}), we propose 
the following scaling law  
\begin{equation}
g(l)= l^2 .
\label{rule}
\end{equation}
With this assumption, the theoretical
prediction of the velocity gradients statistics is given by the following 
equation
\begin{equation}
\label{risultato2}
\psi(\delta v_{l})=\frac{2C}{l^2}\delta v_l K_1 \left(\frac{\delta v_l}{l} \right).
\end{equation}  
We will show that such theoretical prediction is in complete agreement with 
the experimental results.\\
Actually, experimental
data in the dissipation range are difficult to acquire.
We show the comparison between the theoretical prediction
and the experimental data of Van de Water \cite{Wa}. 
These experimental data correspond to the transverse PDF of a turbulent flow
behind a grid in a closed wind tunnel at a nearly-dissipative scale ($\frac{l}{l_d} \sim 10$).
Figure \ref{dist} shows the agreement between the experimental results 
and the theoretical prediction.

Although the moments of the velocity gradients distribution of the 
incoherent turbulence depend on the 
scaling law of the mean temperature $<T_t>$, 
the probabilty $\psi(\delta v_{l})$
has an asymptotic decay close to the exponential function whatever is the 
relation between the scale $l$ and the mean temperature $<T_t>$. 
Indeed, the asymptotic shape of the velocity gradients is given by
\begin{equation}
\psi(\delta v_{l}) \sim e^\frac{-\delta v_l}{c(l)},
\end{equation}
where $c(l)$ is a function of the scale $l$ that depends only on $g(l)$.\\
As shown by Benzi et al. \cite{Be}, the PDF of the velocity gradients 
observed experimentally in the inertial range can be 
obtained by a non-trivial superposition
of stretched exponentials, corresponding to the various exponents of the
multifractal approach. Moreover, such distribution is explicitly
dependent on the Reynolds number.  
On the contrary, at very small scales $l \sim l_d$, the tails of the PDF are
exponential-like and independent of the Reynolds number. 
In practice, at very small scales, the exponential like decay of the PDF is a 
universal feature that, using the present approach, may be described by 
means of ordinary statistical
mechanics.\\
In conclusion, we have presented a description 
of turbulence at very small scales
$l \sim l_d$,
where the coherent structures may be assumed to have been annihilated by 
dissipation.
A thermodynamical model for incoherent turbulence, providing an exact
expression for the PDF (in terms of the modified Bessel function of the
second kind), has been presented. 
We have shown that the velocity gradients fluctuations 
at very small scales may be described  as the momentum {\bf p} of a particle
in a gas with a temperature $T$ that fluctuates slowly. 
Hence, the present model explains how even the random incoherent fluctuations 
in the dissipative range may give rise to non-gaussian statistics.
We hope that this effort will results in a deeper understanding
of the physics underlying the phenomenon of fully developed turbulence
and, more in general, of complex processes characterized by the coexistence
of equilibrium and non-equilibrium conditions.\\
\\
The author is very greatful to G. Buresti for fruitful 
discussions and the continuous help in the preparation of the paper.


\begin{thebibliography}{200}
\bibitem{Ko} A.N. Kolmogorov, Dokl. Akad. SSSR {\bf 30}, 299 (1941) 
\bibitem{Ba} G.K. Batchelor, A.A. Townsend, Proc. Roy. Soc. A {\bf 199}, 238 (1949)
\bibitem{An} F. Anselmet, Y. Gagne, E.J. Hopfinger, R.A. Antonia, J. Fluid Mech. {\bf 140}, 63 (1984)
\bibitem{Vi} A. Vincent, M. Meneguzzi, J. Fluid Mech {\bf 225}, 1 (1991)
\bibitem{Ko2} A.N. Kolmogorov, J. Fluid Mech. {\bf 13}, 82 (1962)
\bibitem{Ob} A.M. Obukhov, J. Fluid Mech. {\bf 13}, 77 (1962)
\bibitem{Sg} E.D. Siggia, J. Fluid Mech. {\bf 107}, 375 (1981)
\bibitem{Kr} R.H. Kraichnan, Phys. Rev. Lett. {\bf 65}, 575 (1990)
\bibitem{Fr} U. Frisch, P.L. Sulem, M. Nelkin, J. Fluid Mech. {\bf 87}, 719 (1978)
\bibitem{Benzi} R. Benzi, G. Paladin, G. Parisi, A. Vulpiani, J. Phys.A {\bf 17}, 3521 (1984)
\bibitem{Me} C. Meneveau, K.R. Sreenivasan, Phys. Rev. Lett. {\bf 59}, 1424 (1987)
\bibitem{She} Z.-S. She, Phys. Rev. Lett. {\bf 66}, 600 (1991)
\bibitem{She2} Z.-S. She, E. Leveque, Phys. Rev. Lett. {\bf 72}, 336 (1994)
\bibitem{Wa} W. van de Water, Physica B {\bf 228}, 185 (1996)
\bibitem{Kra} R.H. Kraichnan, Advanc. Math. {\bf 16}, 305 (1975)
\bibitem{Beck} C. Beck, Phys. Rev. Lett. {\bf 87}, 180601 (2001)
\bibitem{Be} R. Benzi, L. Biferale, G. Paladin, A. Vulpiani, M. Vergassola, Phys. Rev. Lett {\bf 67}, 2299 (1991)

\end{thebibliography}
\end{document}